Raman spectra and electron-phonon coupling in disordered graphene with gate-tunable doping

Isaac Childres[1,2], Luis A. Jauregui[2,3], Yong P. Chen[1,2,3]
[1]*Department of Physics, Purdue University, West Lafayette, IN, 47907, USA*
[2]*Birck Nanotechnology Center, Purdue University, West Lafayette, IN, 47907, USA*
[3]*School of Electrical and Computer Engineering, Purdue University, West Lafayette, IN, 47907, USA*

We report a Raman spectroscopy study of graphene field-effect transistors (GFET) with a controlled amount of defects introduced in graphene by exposure to electron-beam irradiation. Raman spectra are taken at $T$ = 8 K over a range of back gate voltages ($V_g$) for various irradiation dosages ($R_e$). We study effects in the Raman spectra due to $V_g$-induced doping and artificially created disorder at various $R_e$. With moderate disorder (irradiation), the Raman G peak with respect to the graphene carrier density ($n_{FE}$) exhibits a minimum in peak frequency and a maximum in peak width near the charge-neutral point (CNP). These trends are similar to those seen in previous works on pristine graphene and have been attributed to a reduction of electron-phonon coupling strength ($D$) and removal of the Kohn anomaly as the Fermi level moves away from the CNP. We also observe a maximum in $I_{2D}/I_G$ and weak maximum in $I_D/I_G$ near the CNP. All the observed dependences of Raman parameters on $n_{FE}$ weaken at stronger disorder (higher $R_e$), implying that disorder causes a reduction of $D$ as well. Our findings are valuable for understanding Raman spectra and electron-phonon physics in doped and disordered graphene.

Introduction

Graphene has received much attention in the scientific community because of its distinct properties and potentials in nanoelectronic applications.[1, 2] Raman spectroscopy,[3-5] which identifies vibrational modes using only laser excitation, is a powerful, non-invasive method to measure many important characteristics of graphene,[6] such as layer number, defect density and carrier concentration.

In graphene, the Stokes phonon energy shift of laser excitation creates two main peaks in the Raman spectrum. The G peak (~1580 cm$^{-1}$) is the primary in-plane vibrational mode, caused by the $E_{2g}$ phonon at the Γ point. The other major peak in graphene is 2D (~2690 cm$^{-1}$), which is created by a process of double scattering of $A_{1g}$ phonons with an electron-hole pair between K to K'.[4]

The phonons responsible for the G mode have strong electron-phonon interactions, resulting in Kohn anomalies [7] in the phonon dispersion, which soften phonons at wavevector $q \sim 2k_F$.[8] Doping in graphene, which shifts the Fermi level ($E_F \propto k_F$) away from the Dirac point, moves the Kohn anomaly (located at $2k_F$) away from $q = 0$, where the G mode originates. This causes a stiffening of the G peak, increasing phonon energy.[8,9] Increased doping also sharpens Raman peaks by reducing electron-phonon interactions through the blockage of decay channels from phonons into electron-hole pairs.[8,10] In addition, increased carrier doping in graphene has also been shown to decrease the intensity of the 2D peak.[11] By examining the widths, frequencies and intensities of the G and 2D peaks in a graphene sample, one can gain information about its layer number, doping and electron-phonon coupling strength ($D$).

Another well-studied peak in the Raman spectrum of graphene is the D peak (~1350 cm$^{-1}$), which is not activated in pristine graphene because of crystal symmetries. In order for the D peak to occur, a charge carrier must be excited and inelastically scattered by a phonon, then elastically scattered by a lattice defect or grain boundary to recombine.[12] Raman spectroscopy is one of the most widely used methods of defect characterization due to the strong dependence of graphene's Raman D peak on disorder. Disorder in graphene not only activates the D peak, which is caused by scattering from K to K' (intervalley), but also gives rise to the D' peak (~1620 cm$^{-1}$), caused by scattering from K to K (intravalley), and D+D' (~2940 cm$^{-1}$), a combination scattering peak.[12] As has been previously reported, one can use the ratio of Raman peak intensities ($I_D/I_G$) to characterize the level of disorder in graphene.[13-19] As disorder in graphene increases, $I_D/I_G$ displays 2 different behaviors: a regime of "low defect density," where $I_D/I_G$ will increase with increased disorder as increasing defect density creates more elastic scattering; and a regime of "high defect density," where $I_D/I_G$ will decrease with increased disorder as an increasing defect density results in a more amorphous carbon structure, attenuating all Raman peaks.[12]

There are very few studies, however, examining the effect of graphene carrier density ($n_{FE}$) on the Raman peaks in *disordered* graphene. Such a study will be important for gaining a more complete understanding of phonons and electron-phonon coupling in disordered graphene. In this work, we directly investigate the dependence of graphene's Raman characteristics on both $n_{FE}$ and the level of disorder in a graphene sample with disorder created by electron-beam irradiation.

Methods

Our graphene samples are fabricated using a similar method as in our previous publications.[16, 17] We perform micromechanical exfoliation [2] of highly ordered pyrolytic graphite (HOPG, "ZYA" grade, Momentive Performance Materials) onto a p-doped Si wafer with 300 nm of $SiO_2$. Single-layer graphene flakes, typically around 100 $\mu m^2$ in size, are identified using color contrast with an optical microscope [20] and then confirmed with Raman spectroscopy.[12] Graphene field-effect devices are then fabricated using electron-beam lithography. The electrical contacts (5 nm Cr/35 nm Au) are fabricated by electron-beam evaporation.

The graphene sample is then placed in a scanning electron microscope (SEM), and a 25 μm by 25 μm area is continuously scanned by the electron beam to create disorder, as in our previous work.[16] The beam's kinetic energy is 30 keV, and the beam current is tuned so that the exposure takes 60 seconds of scanning. For instance, if the target irradiation dosage were 300 e$^-$/nm$^2$, a current of 0.4 nA would be used. In addition, the same sample is irradiated multiple times to reach a total accumulated dosage ($R_e$).

For instance, after measuring the device at $R_e$ = 300 e⁻/nm², it is irradiated with a further 700 e⁻/nm² (0.933 nA for 60 seconds) to arrive at $R_e$ = 1000 e⁻/nm². We note that the efficacy for the electron beam to create defects in graphene can vary for different experiments and $R_e$ is related to, but does not provide a quantitative measurement of the defect length ($L_D$). All data shown in this paper are from a single graphene device and were taken over the course of a few days, though we have measured similar behaviors in several other samples.

After each successive exposure, the graphene device is removed from the SEM and transferred to a microscopy cryostat (Cryo Industries RC 102-CM) with electrical connections and an optical window and then brought to a temperature of ~8 K and a vacuum pressure of ~$10^{-5}$ mTorr. Field effect measurements (resistance versus back gate, $V_g$) are performed to determine capacitively induced $n_{FE}$ of the graphene using

$$n_{FE} = \frac{\varepsilon_0 \varepsilon (V_g - V_D)}{te}, \qquad (1)$$

where $\varepsilon_0$ and $\varepsilon$ are the permittivities of free space and SiO$_2$ respectively, $t$ is the thickness of the SiO$_2$ substrate, $e$ is the electron charge, $V_D$ is the back gate voltage corresponding to the charge neutral point (CNP).

Raman spectroscopy is performed using a confocal microscope system (Horiba Xplora) with an excitation laser of 532 nm at a power of 0.1 mW incident on graphene, with each spectrum presented as an average of 3 measurements of 20 seconds each. Using a 100X objective, the Raman laser spot size is smaller than 1 µm². We characterize each Raman peak (G, D and 2D) by a Lorentzian fit,

$$f(\omega) = \frac{1}{2\pi} \cdot \frac{I\Gamma}{(\omega - \omega_0)^2 - \left(\Gamma/2\right)^2}, \qquad (2)$$

where $\omega_0$ is the peak position, $\Gamma$ is the full width at half max (FWHM) and $I$ is the integrated intensity of the full peak curve. Near 1600 cm$^{-1}$ in the Raman spectrum for a disordered sample there is an overlap of the G and D' peaks, and we fit those peaks together using a double-Lorentzian fit.

Data

Fig. 1 shows the Raman spectra for our graphene device at various $R_e$ and $n_{FE}$. Fig. 1a shows representative spectra from $R_e$ = 0 e$^-$/nm$^2$ to 70000 e$^-$/nm$^2$ at the graphene device's CNP. The spectra progression from unirradiated to highly irradiated ($R_e$ = 30000 e$^-$/nm$^2$) shows a trend of decreasing 2D intensity ($I_{2D}$) and increasing D, D' and D+D' intensities with increasing irradiation. Fig. 1b and 1c show the spectra near the G and 2D peaks, respectively, for the unirradiated device at different $V_g$ ranging from -60 V to +60 V away from the CNP. This progression of spectra shows a minimum in the G peak frequency ($\omega_G$) and a maximum in the G peak width ($\Gamma_G$) near the device's CNP. This is consistent with previous studies of G peak dependence on $n_{FE}$ for pristine graphene (with no appreciable disorder to have an observable D peak).[8-11] Fig. 1d-f show the spectra near the D, G and 2D peaks, respectively, for the same device after moderate irradiation ($R_e$ = 3000 e$^-$/nm$^2$) at $V_g$ ranging from -40 to +60 V away from the CNP. Again we see a trend of decreased $\omega_G$ and increased $\Gamma_G$ as $V_g$ approaches the CNP. These trends can be seen more clearly in Fig. 2. We also note that the field-effect measurements show a trend of decreasing carrier mobility and minimum conductivity as irradiation increases, which is consistent with our previous report.[16]

Fig. 2 shows the extracted $\omega_G$ (a), $\Gamma_G$ (b) and G peak intensity ($I_G$, c) as a function of $V_g$-$V_D$, which is proportional to $n_{FE}$ (top axis). In addition to a minimum in $\omega_G$ near the CNP for low- to medium-levels of irradiation, we also see a peak in $\Gamma_G$ near the CNP for the same range of irradiation. However $I_G$ shows no significant dependence on $n_{FE}$ from -4*10$^{12}$ cm$^{-2}$ to 4*10$^{12}$ cm$^{-2}$ for a fixed $R_e$, nor on irradiation up to $R_e$ = 3000 e⁻/nm². For higher $R_e$, the G peak becomes significantly wider and the overall intensity increases. In addition, at these high irradiation dosages ($R_e$ = 30000 e⁻/nm² and 70000 e⁻/nm²), $\omega_G$ and $\Gamma_G$ show very weak dependence on $n_{FE}$ within the resolution of the experiment.

We note that the maxima and minima in Fig. 2 do not occur exactly at the CNP, but at some smaller $V_g$. Similar features can also be seen in other figures. We believe this is due to the effects of local, laser-induced doping.[21] We also note the extracted Raman parameters can show fluctuation (nonrepeatable) at larger $R_e$, where we expect more spatial inhomogeneity of $n_{FE}$ due to charge puddles caused by irradiation. The fluctuation may be caused by small variations in the location of the Raman laser spot, which can be caused by small variations in the temperature in the cryostat.

Next we look at the effect of $n_{FE}$ on the Raman 2D peak for different $R_e$. We see no clear dependence of the 2D FWHM ($\Gamma_{2D}$) on $n_{FE}$, however the 2D peak frequency ($\omega_{2D}$) has a broad, weak minimum near the CNP at low irradiation dosages. In Fig. 3c, we see a maximum $I_{2D}$ near the CNP up to $R_e$ = 3000 e⁻/nm². We also see a decrease in the overall intensity of the 2D peak with increasing irradiation. Due to the maximum in $I_{2D}$ near the CNP, when we plot the ratio of the 2D and G peak intensities ($I_{2D}/I_G$) as a function of $V_g$ (Fig. 3d), we see a clear maximum of $I_{2D}/I_G$ near the CNP for irradiation dosages up to $R_e$ = 1000 e⁻/nm². $I_{2D}/I_G$ decreases with increased irradiation, and at higher irradiation dosages ($R_e$ = 10000 e⁻/nm² and 30000 e⁻/nm²), its dependence on $n_{FE}$ completely disappears.

Finally we look at the effect of $n_{FE}$ and disorder on the D peak of graphene. The D peak shows no clear dependence of peak frequency ($\omega_D$) or FWHM ($\Gamma_D$) on carrier density as observed. In Fig. 4c we see a very weak, broad peak in the D peak intensity ($I_D$) near the CNP for $R_e$ = 300 e⁻/nm² and 1000 e⁻/nm² as well as an increase in $I_D$ for all $n_{FE}$ as the irradiation dosage increases up to $R_e$ = 30000 e⁻/nm². We also plot the intensity ratio $I_D/I_G$ in Fig. 4d, where we can see a weak, broad peak near the CNP for $R_e$ = 300 e⁻

/nm$^2$ and 1000 e$^-$/nm$^2$. Fig. 4d also shows a clear trend of increasing $I_D/I_G$ with increasing irradiation up to $R_e$ = 10000 e$^-$/nm$^2$, which is expected.

From Fig. 2-4, we also note an overall decrease in the frequency and an increase in the FWHM for the D, G and 2D peaks with increasing $R_e$. This can be seen more clearly in Fig. 5a and b, which plot the change in the D, G and 2D peak frequency and FWHM at $V_g$-$V_D$ = 0 from an unirradiated (G and 2D peaks) or lightly irradiated (D peak) state as a function of $R_e$, which we believe to be proportional to the defect density (where the proportionality constant depends on the details of the electron beam interaction with respect to the graphene, which are not known). We see clear trends of decreasing frequency and increasing FWHM for all peaks as the defect density increases (increased irradiation). Of the three peaks plotted, the 2D peak shows the largest change in frequency and FWHM and shows the strongest dependence on $R_e$, probably due to the fact that 2D is a double-phonon peak.[19]

Analysis

The trends of increasing $\omega_G$ and decreasing $\Gamma_G$ for increasing $n_{FE}$ we see in Fig. 2 at lower $R_e$ are similar to previous reports for pristine graphene.[8-11] These trends are attributed to the removal of the Kohn anomaly and decreased electron-phonon coupling for increased $n_{FE}$. Our results show that such mechanisms still exist in moderately disordered graphene. On the other hand, for $R_e$ < 10000 e$^-$/nm$^2$, we observe that $I_G$ does not vary appreciably with either $R_e$ or $n_{FE}$ within our measurement range.

As disorder increases (increasing $R_e$), $\omega_G$ and $\Gamma_G$ show less dependence on $n_{FE}$. This could be caused by disorder dominating the phonon scattering processes, therefore reducing the effect of electron-phonon coupling. We can calculate the electron-phonon coupling strength ($D$) for different $R_e$ using a linear approximation with time-dependent perturbation theory [9]

$$\hbar\omega_G - \hbar\omega_G^0 = \frac{A_{uc}D^2}{2\pi\hbar\omega_G M v_F^2}|E_F|, \quad \text{where} \quad E_F = \hbar v_F\sqrt{\pi n_{FE}}, \tag{3}$$

$\omega_G^0$ is the G peak frequency at $E_F = 0$ (CNP), $A_{uc} = 0.51$ nm$^2$ is the area of the graphene unit cell, $M = 2*10^{-26}$ kg is the mass of a carbon atom and $v_F = 10^6$ m/s is the Fermi velocity in graphene. This equation can be used sufficiently far away from the Dirac point where the trend of G peak energy ($E_G = \hbar\omega_G$) versus Fermi energy ($E_F$) is approximately linear. We perform this fitting in Fig. 5c (solid lines), which plots $E_G$ versus $E_F$ for different $R_e$. We plot the extracted $D$ as a function of $R_e$ in Fig. 5d (and $D$ as a function of $I_D/I_G$ in the inset). We see that $D$ decreases with increasing $R_e$ (stronger disorder with higher $I_D/I_G$). For the unirradiated sample we find $D$ = 14.7 ev/Å, which agrees fairly well with previous works.[9] $D$ then decreases to ~7 eV/Å for $R_e$ = 30000 e$^-$/nm$^2$.

Another way to extract $D$, as also discussed in ref. 9, is to use the total change in $\Gamma_G$ between the CNP and sufficiently high $n_{FE}$[9]

$$\Delta\Gamma = \frac{A_{uc}D^2}{8Mv_F^2} \tag{4}$$

We also plot the extracted values of $D$ from the data in Fig. 2b based on this equation as a function of $R_e$ (except for $R_e$ = 30000 e$^-$/nm$^2$ where the fluctuation in $\Gamma_G$ is too large to allow such analysis) in Fig. 5d (and $D$ as a function of $I_D/I_G$ in the inset), and find the values in general agreement with $D$ calculated from the peak frequency data (Eq. 3), with $D$ = 15.3 eV/Å for unirradiated graphene, and $D$ decreasing for larger $R_e$, again suggesting that increasing disorder weakens electron-phonon coupling.

We note the possibility that increased charge inhomogeneity at larger $R_e$ could cause a decreased dependence on $V_g$-induced $n_{FE}$. However, from our field-effect data we conclude that the inhomogeneity is on the order of < $2*10^{12}$ cm$^{-2}$ at the highest $R_e$ measured. This is significantly smaller than our measurement range, implying inhomogeneity alone is not the cause of the disappearance of $n_{FE}$ dependence in $\omega_G$ and $\Gamma_G$.

We also see trends of decreasing peak position and increasing FWHM with increasing $R_e$ for the G, 2D and D peaks, which can be seen clearly in Fig. 5 and are consistent with the results in Ref. 19. We attribute the trend of decreasing frequency to a softening of the lattice caused by defects, which would reduce the energy of lattice vibrational modes. We fit the peak frequency trends to a phenomenological power law, $\Delta\omega \propto R_e^p$, and find power dependences of p = 0.88, 0.19 and 0.56 for the D, G and 2D peaks respectively. We can attribute the increasing FWHM to increased phonon scattering due to defects, which will decrease phonon lifetime. In fact, the FWHM can be described as a sum of contributions from phonon-phonon interactions ($\gamma^{gn}$), electron-phonon interactions ($\gamma^{EPC}$)[12] and phonon defect scattering ($\gamma^{D}$). We have demonstrated a decrease in $\gamma^{EPC}$ with increased disorder and would expect a similar decrease in $\gamma^{gn}$. These reduced interactions would reduce the peak FWHM, so an increased FWHM with increased disorder must be caused by increased $\gamma^{D}$. One other trend to note is that $\Gamma_G$ remains relatively constant for low levels of disorder, which is consistent with previous results.[22]

The 2D and D peak frequencies and widths don't have a significant dependence on $n_{FE}$, however their integrated intensities show some dependence on $n_{FE}$ at $R_e$ < 3000 e$^-$/nm$^2$. At these dosages, both $I_{2D}$ and $I_D$ decrease with increased $n_{FE}$. For the 2D peak, this dependence has been previously studied in pristine graphene and the intensity ratio $I_{2D}/I_G$ cited as an important parameter to estimate doping concentration [10] (in addition, $I_{2D}/I_G$ is commonly used to determine the number of layers in graphene),[23] however we note that attention should also be paid to the disorder level, as increased $R_e$ causes a weakening of $I_{2D}/I_G$'s dependence on $n_{FE}$ and an overall decrease in its value. In addition, the strong dependence of $I_D$ on disorder has been used to characterize $L_D$ in terms of the intensity ratio $I_D/I_G$.[13-18] At low values of irradiation, however, we show this ratio also has a weak dependence on $n_{FE}$, and this dependence has not been captured in previous models of $L_D$ with respect to $I_D/I_G$.

We have demonstrated that both disorder and $n_{FE}$ affect a number of Raman peak parameters, including peak position, width and intensity for the D, G and 2D modes. We measured these effects and have concluded that increased $n_{FE}$ in graphene causes the removal of the Kohn anomaly and decreases phonon scattering, while increased disorder reduces electron-phonon coupling and increases phonon scattering. Our results are valuable for understanding Raman spectra and electron-phonon physics in doped and disordered graphene, and they suggest attention should be paid to both disorder and carrier density when characterizing graphene through Raman spectra.

Correspondence should be directed to: *ichildre@purdue.edu*

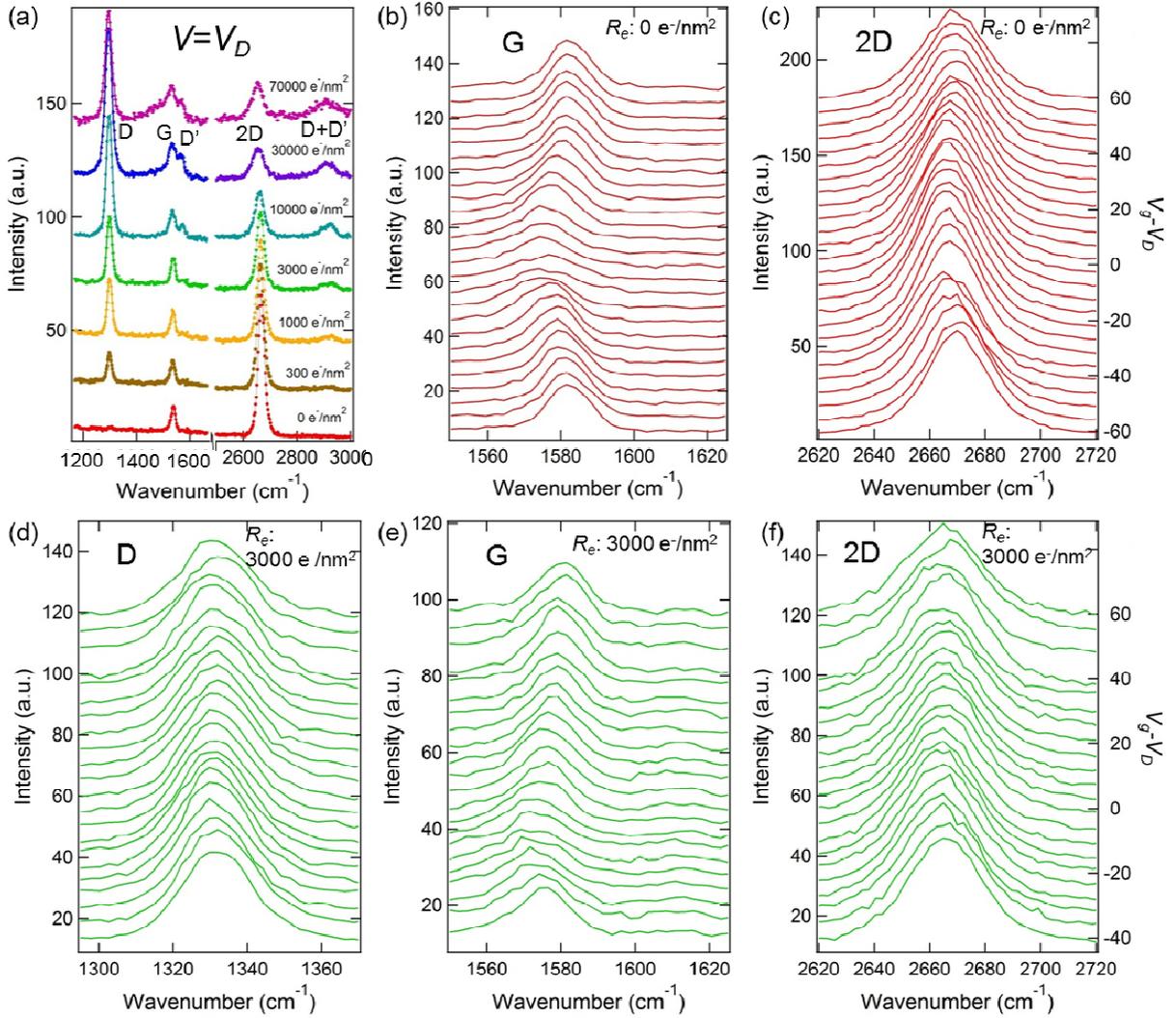

**Figure 1**: (a) Raman spectra (measured with a 532 nm excitation laser) of graphene at its charge-neutral point ($V_D$) for different dosages ($R_e$) of irradiation by a 30 keV electron beam. Representative Raman peaks are labeled in the full spectrum for $R_e$ = 30000 e$^-$/nm$^2$. (b, c) The G and 2D peaks, respectively, for unirradiated graphene at a range of back gate voltages ($V_g$, plotted on the right axis of (c) relative to $V_D$). (d, e, f) Raman spectra of the D, G, and 2D peaks, respectively, at different $V_g$ (shown on the right axis of (f)) for the same graphene sample with an irradiation dosage $R_e$ = 3000 e$^-$/nm$^2$. The spectra of all plots have been offset vertically for clarity.

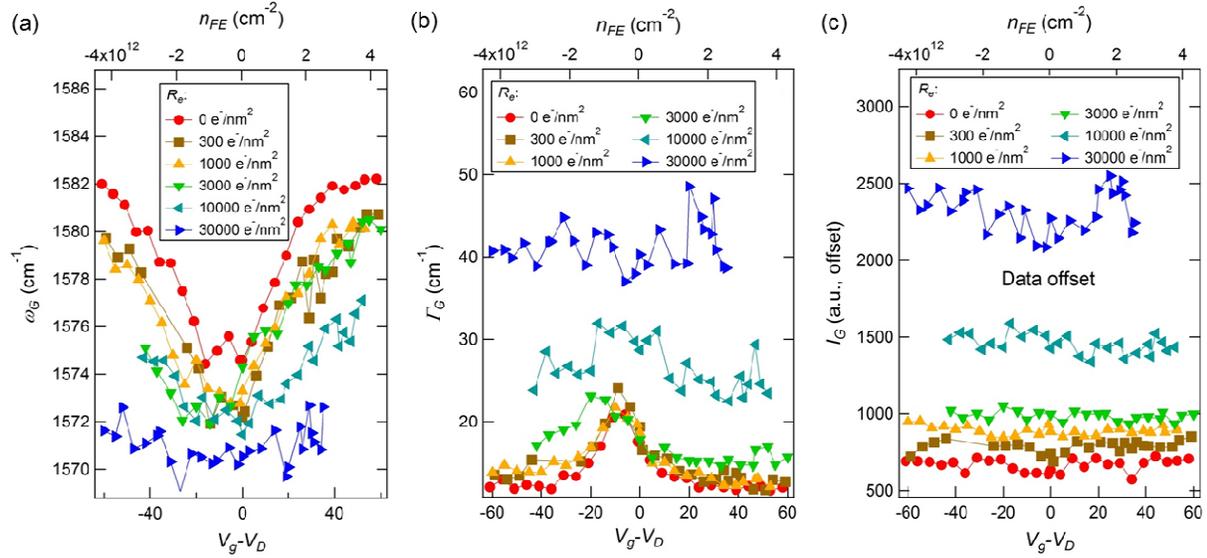

**Figure 2**: Raman G peak frequency (a, $\omega_G$), FWHM (b, $\Gamma_G$) and integrated intensity (c, $I_G$) plotted against the gate voltage relative to the Dirac point, $V_g$-$V_D$ (proportional to the carrier density, plotted on the top axis), for different dosages, $R_e$, of irradiation. In (c), the data sets for $R_e$ > 0 e$^-$/nm$^2$ are offset consecutively by 100 vertically for clarity.

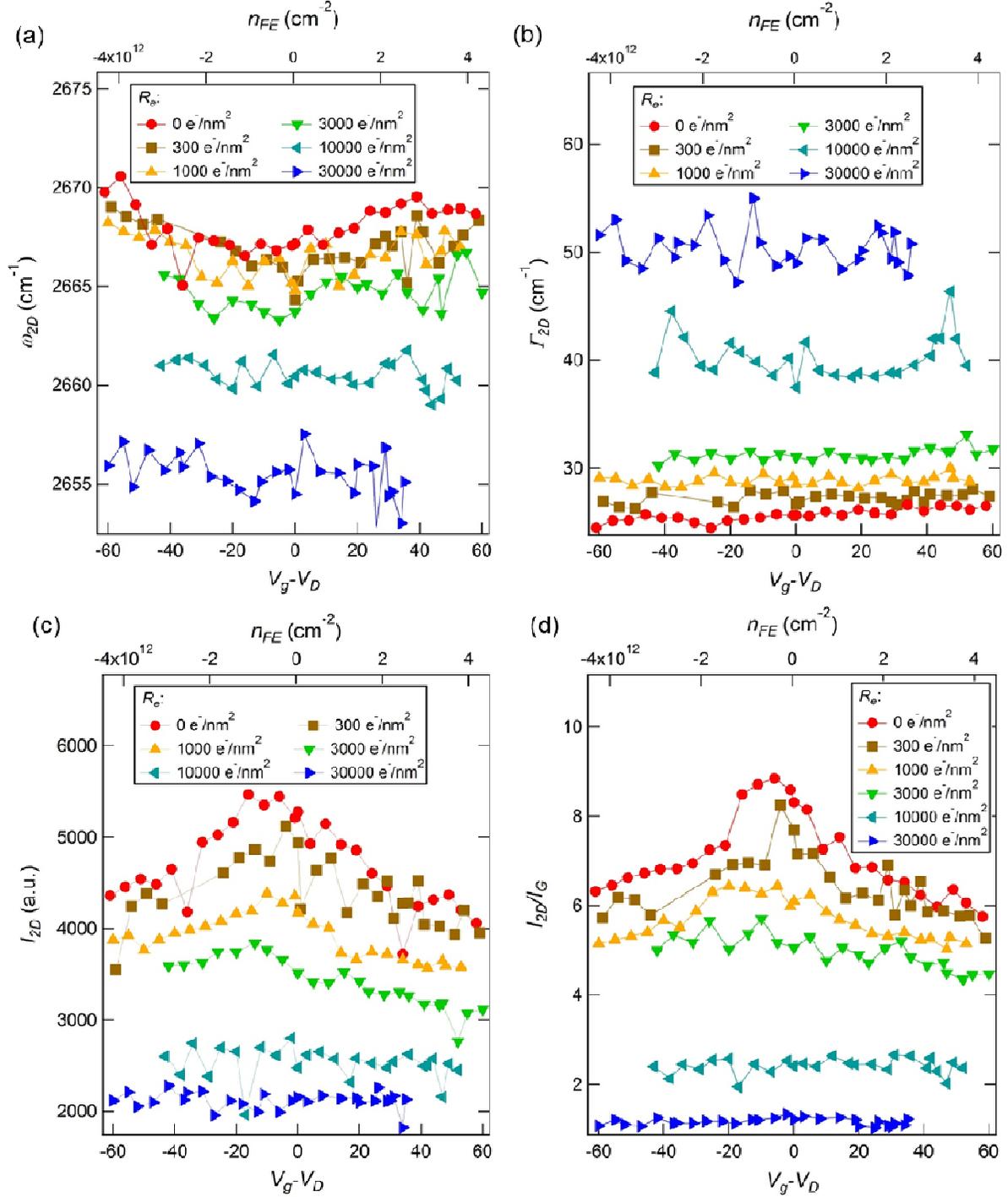

**Figure 3**: Raman 2D peak frequency (a, $\omega_{2D}$), FWHM (b, $\Gamma_{2D}$), integrated intensity (c, $I_{2D}$) and intensity ratio (d, $I_{2D}/I_G$) plotted against the gate voltage shift relative to the Dirac point, $V_g$-$V_D$ (proportional to the carrier density, plotted on the top axis), for different dosages, $R_e$, of irradiation.

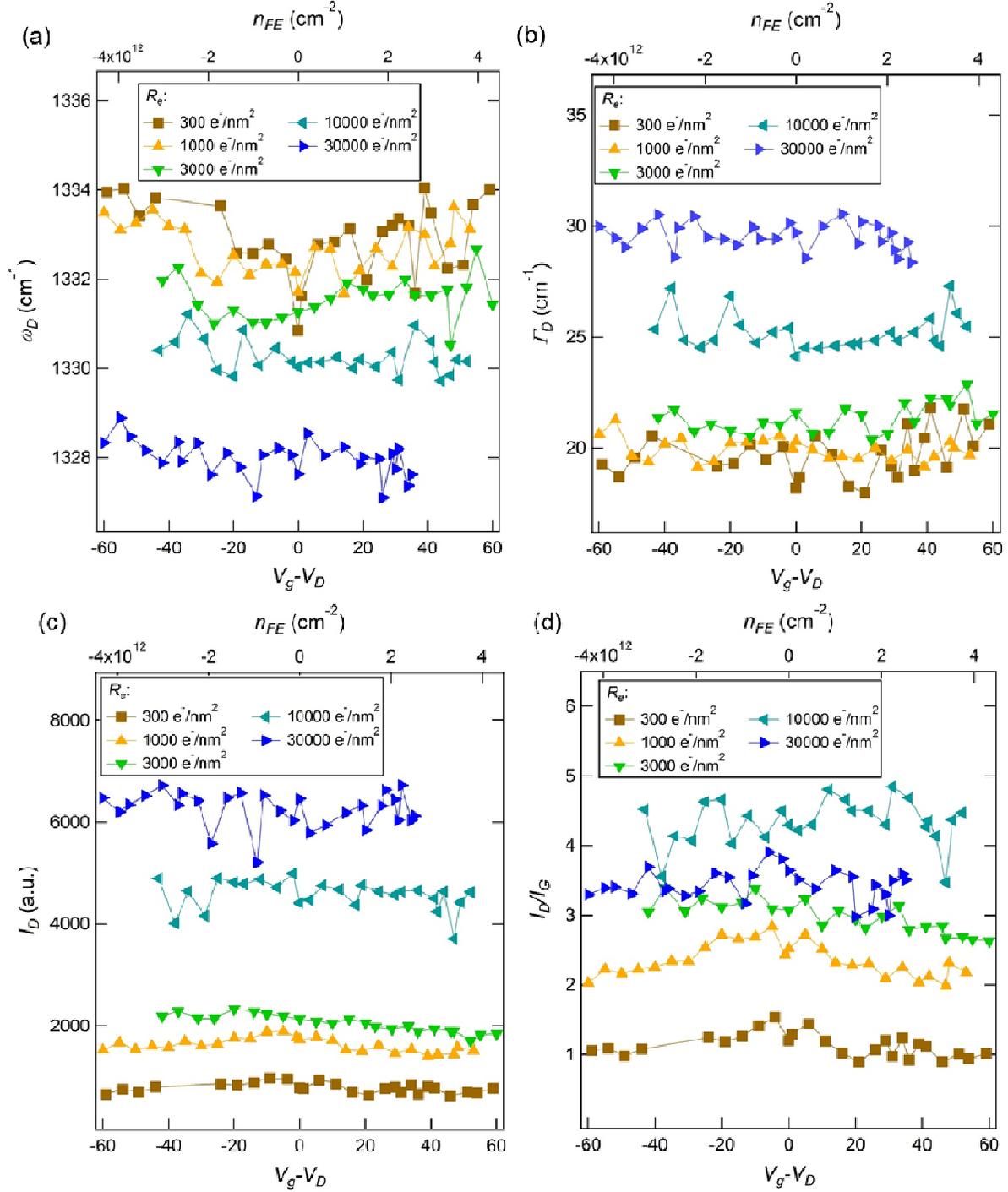

**Figure 4**: Raman D peak frequency (a, $\omega_D$), FWHM (b, $\Gamma_D$), integrated intensity (c, $I_D$) and intensity ratio (d, $I_D/I_G$) plotted against the gate voltage shift relative to the Dirac point, $V_g$-$V_D$ (proportional to the carrier density, plotted on the top axis), for different dosages, $R_e$, of irradiation.

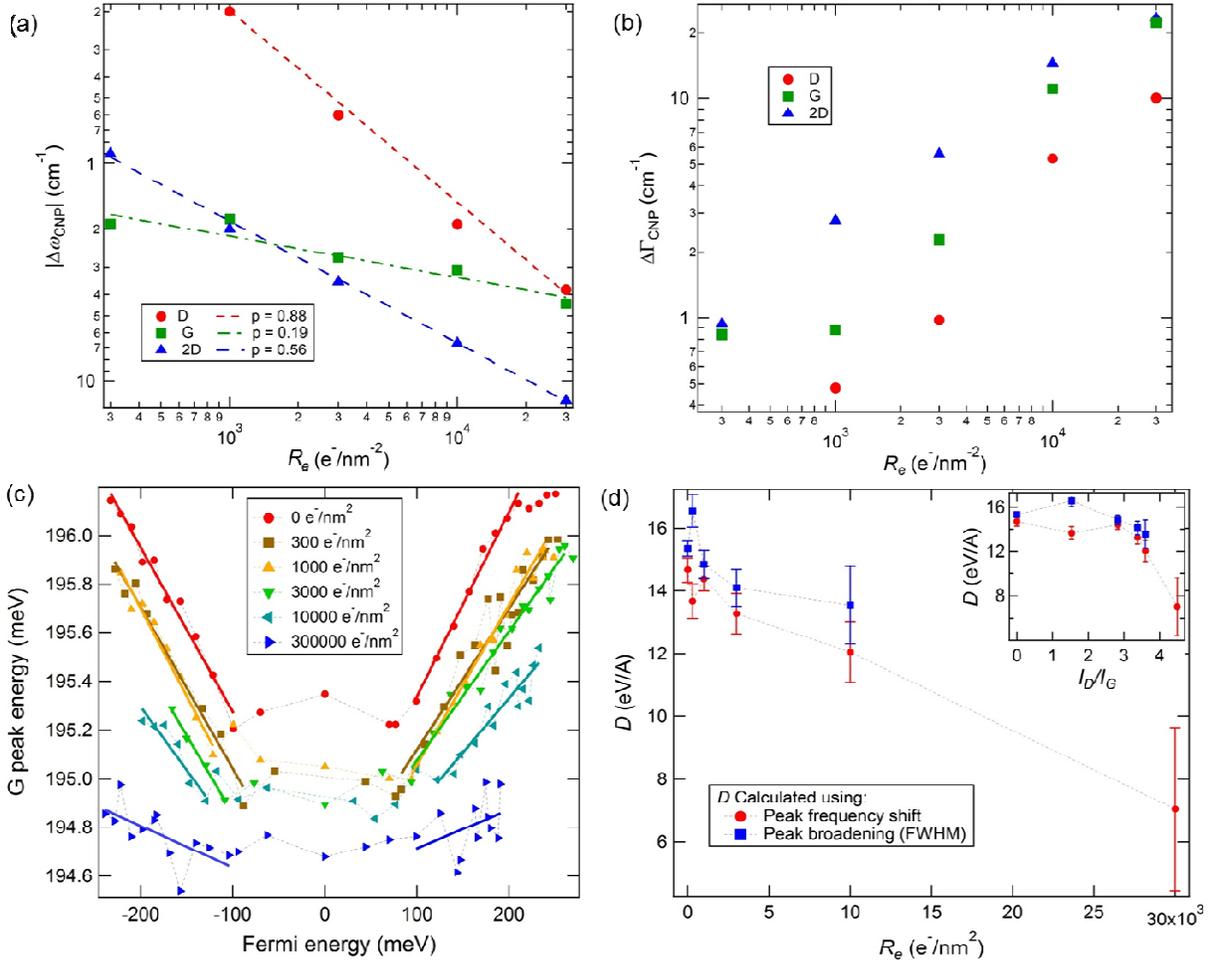

**Figure 5:** The change of peak frequency, (a, $\Delta\omega$) and FWHM, (b, $\Delta\Gamma$) from an unirradiated state for the Raman D, G and 2D peaks at the CNP ($V$-$V_D$ = 0) versus the irradiation dosage, $R_e$, plotted in a log-log scale. The dashed lines in (a) are power law fittings to y $\propto$ $R_e^p$. Since we see no significant D peak in the unirradiated state, $\Delta\omega$ and $\Delta$FWHM for the D peak are plotted relative to $R_e$ = 300 e$^-$/nm$^2$. (c) The energy of the Raman G peak ($E_G$) versus the Fermi energy ($E_F$) of the graphene for different dosages, $R_e$, of irradiation. The solid lines are linear fittings (Eq. 3) far away from $E_F$ = 0. (d) Electron-phonon coupling strength ($D$) versus $R_e$. For each $R_e$, $D$ is calculated both from the measured $E_F$ dependence of $\omega_G$ (data in c) by fitting to Eq. 3 as well as from the $E_F$ dependence of $\Gamma_G$ (broadening near CNP, from FWHM data in Fig. 2b) using Eq. 4.

16